\documentclass{mem}
\usepackage{natbib}\usepackage{txfonts}
\usepackage{graphicx}
\usepackage[a4paper]{hyperref}
\idline{75}{282}
\begin{document}
\def\teff{$T\rm_{eff }$}

\title{Synthesis of the Beryllium 3131$\AA$ Spectral Region}

   \subtitle{}

\author{Johanna F. \,Ashwell\inst{1}
\and R.D. \, Jeffries\inst{1}
\and Barry \, Smalley\inst{1}}

  \offprints{Johanna F. Ashwell}

\institute{Astrophysics Group, School of Chemistry and Physics, Keele University,
Staffordshire, ST5 5BG, UK
\email{jfa@astro.keele.ac.uk}}

\authorrunning{Ashwell}

\titlerunning{Synthesis of the Beryllium 3131$\AA$ Spectral Region}

\abstract{
The Beryllium spectral region of the Sun, Procyon and 4 stars in the open cluster NGC6633 up to
\teff = 7500K have been synthesised using {\sc atlas}9 model atmospheres and the {\sc moog} spectral synthesis program.

The line list used for these syntheses has been modified from the {\sc atlas}9 line list to improve the quality of the fits in light of the improved opacities in the new version of the {\sc moog} code.

Significant changes have been made to the Mn {\sc i} line at {\sc atlas}9 wavelength 3131.037$\AA$ and an OH line has been added at 3131.358$\AA$. In addition there are a number of minor changes to {\it gf}-values throughout the synthesised region thus improving the fit for the spectra across the temperature range considered.

}

\maketitle{}

\section{Introduction}
\label{intro}
The synthesis of the Beryllium region presented in this paper was carried out with the objective of determining the Be abundance of 4 stars in the open cluster NGC6633 (one of which exhibits chemical peculiarities) using new CCD data from the UV Visual Echelle Spectrograph (UVES) on the Very Large Telescope (VLT) (see \citet{ashwell05}).

As both components of the Be {\sc ii} doublet are blended with other lines it is necessary to synthesis this region of the spectrum. This synthesis has been carried out using an updated version of the LTE analysis code {\sc moog} \citep{sneden02} and atmospheres interpolated for the individual stellar parameters from the Kurucz {\sc atlas}9 grids.

\section{Line List Selection}
\label{llselection}
The 3131$\AA$ spectral region which contains the Be {\sc ii} resonance doublet is rich with strong atomic and molecular lines in solar type stars. This results in substantial line absorption and a deficit of true continuum regions making normalisation difficult.

Though, spectrum synthesis can reduce normalisation problems by taking into account possible blending features, laboratory studies of the identification, precise wavelengths and oscillator strengths of many features in this spectral region are limited.

The initial line list used consists of selected atomic and molecular lines from the {\sc atlas}9/Kurucz CDROMs with the lines selected on the basis of having an excitation potential less than 10.0 eV and log {\it gf} greater than 10.0.

\section{Line List Calibration}
\label{llcalibration}
\begin{figure}[t!]
\resizebox{\hsize}{!}{\includegraphics[clip=true]{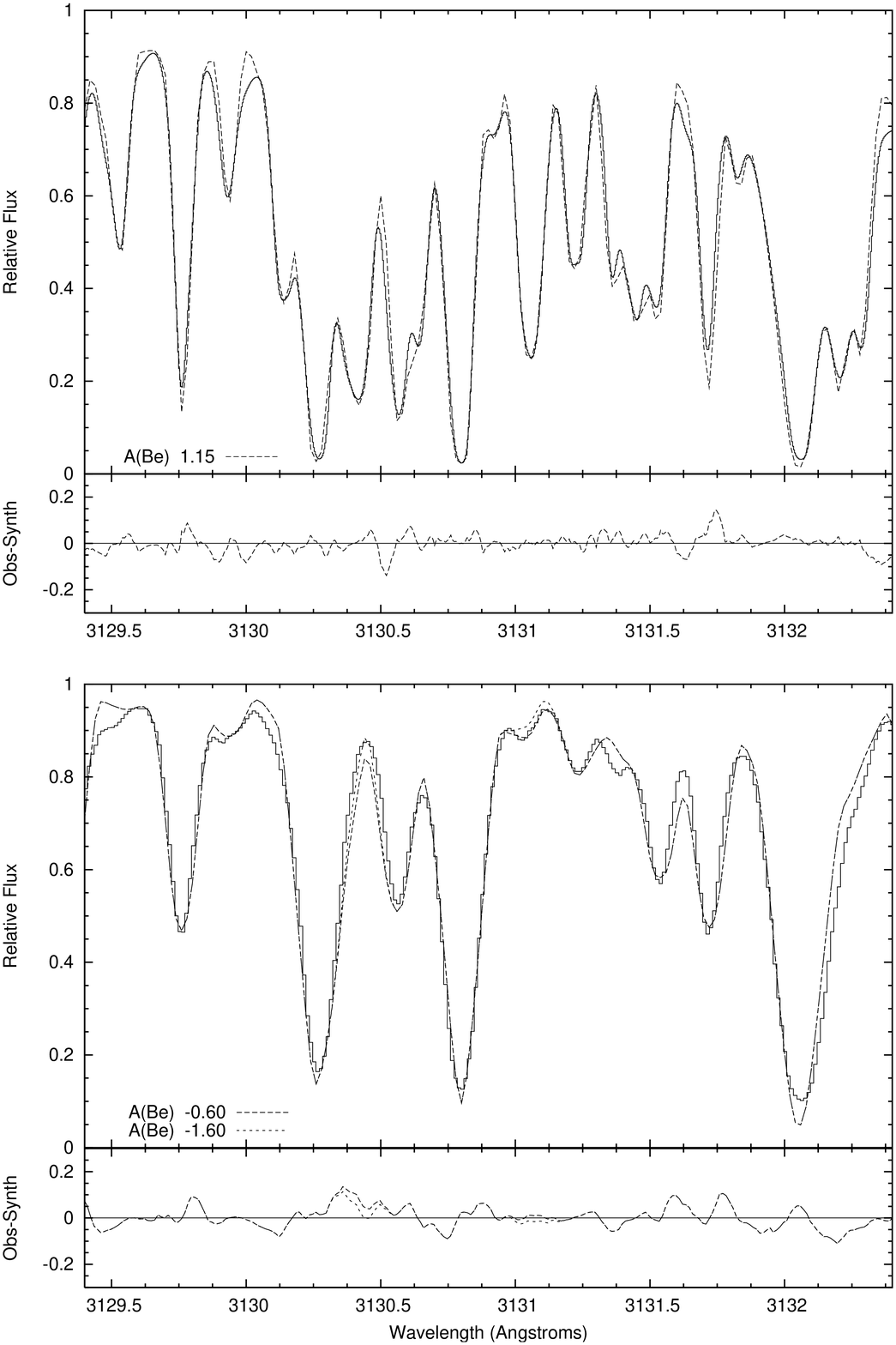}}
\caption{\footnotesize
Spectral synthesis of the Solar Be {\sc ii} doublet (3130.4$\AA$ and 3131.1$\AA$) region from the NOAO Solar atlas (upper) and Procyon (lower). The upper portion of each plot shows the spectrum (solid line) and synthesis (dashed line). The lower portion of each plot shows the residuals.}
\label{solarprocyon}
\end{figure}

Throughout the 3129.5 - 3132.5$\AA$ region synthesised the oscillator strengths for a number of lines were altered from their {\sc atlas}9 values to better fit both the NOAO Solar Atlas \citep{kurucz84} {\it and} a spectrum of Procyon (F5 IV) obtained with UVES by \citet{bagnulo03} (see Figure \ref{solarprocyon}).

Possible arbitrary changes were avoided by adopting the philosophy that {\it gf}-values which needed the least adjustment should be chosen where a number of lines affected the same feature. The resulting solar abundance, A(Be) = 1.15, is in excellent agreement with the photospheric abundance quoted in \citet{anders89}, whilst Procyon is confirmed as very Be-depleted, A(Be)$\leq$-0.5.

To simultaneously match the Procyon spectrum, with \teff = 6700K, log g = 4.05 \citep{lemke93}; and the Sun a few major changes were made.

\section{Line List Modifications}
\label{llmodifications}
Of these major changes made to the {\sc atlas}9 line list only two affect the Be {\sc ii} lines directly. The first is an alteration of both the wavelength and {\it gf}-value of an Mn {\sc i} line and the second is the addition an OH molecular line where a line is clearly missing in the {\sc atlas}9 line list.

\begin{figure*}[]
\resizebox{\hsize}{!}{\includegraphics[clip=true]{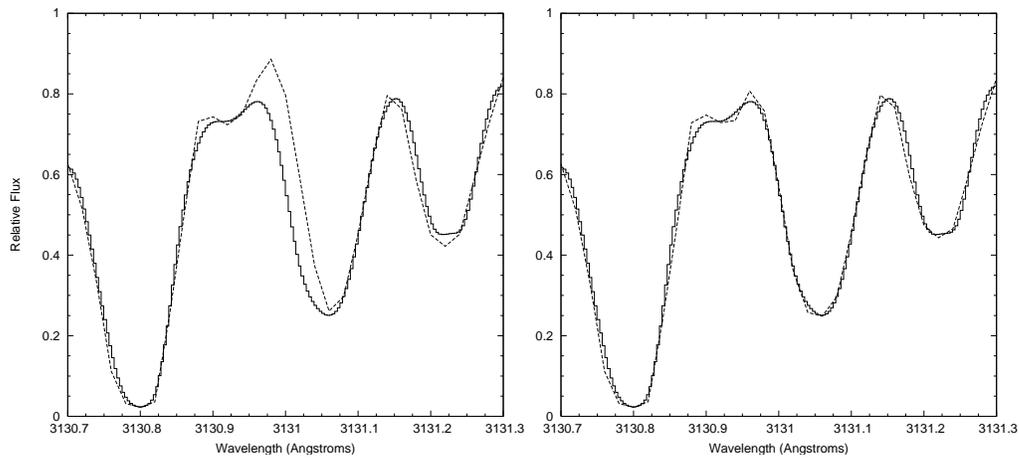}}
\caption{\footnotesize
A magnified section of Figure \ref{solarprocyon} showing the changes made to the Mn {\sc i} line at 3131.037$\AA$. The left plot shows the original {\sc atlas}9 line list parameters and the right plot shows the fit achieved by altering these parameters (see section \ref{llmodifications}).}
\label{mnline}
\end{figure*}

The wavelength of the Mn {\sc i} in the {\sc atlas}9 list is stated as 3131.037$\AA$ with a log {\it gf}-value of 1.725 and gives the fit shown in the left plot of Figure \ref{mnline} using {\sc moog}. The synthesis (dashed line) is clearly discrepant from the spectrum (solid line) appearing to be positioned at too longer wavelength.

However, a remedy to this problem is suggested by \citet{king97}. They suggest that shifting the Mn {\sc i} line to 3131.017$\AA$ (a shift of ~0.02$\AA$ is not unreasonable when considering errors) could account for the discrepancy. Therefore, by shifting this line and increasing its {\it gf}-value by +1.56 dex the fit shown in the right hand plot of Figure \ref{mnline} is achieved.

\begin{figure*}[]
\resizebox{\hsize}{!}{\includegraphics[clip=true]{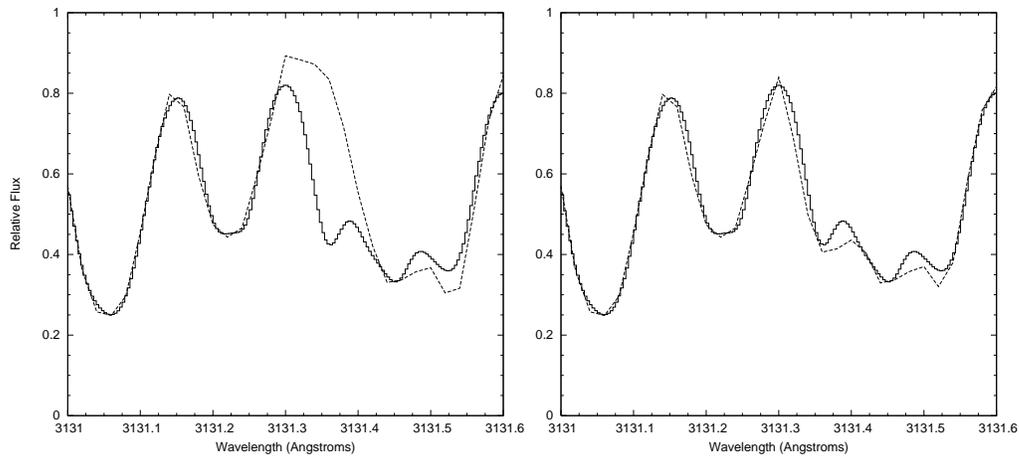}}
\caption{\footnotesize
A magnified section of Figure \ref{solarprocyon} showing the changes made redward of the Be {\sc ii} doublet. The left plot shows the original {\sc atlas}9 line list parameters and the right plot shows the fit achieved by adding an OH molecular line (see section \ref{llmodifications}).}
\label{redbeline}
\end{figure*}

The addition of an OH molecular line as seen in the right hand plot of figure \ref{redbeline} corrects for the apparently missing line at 3131.350$\AA$. The added line with a wavelength of 3131.358$\AA$, excitation potential of 1.941 and log {\it gf}-value 1.347 was used in the line list for various King papers including \citet{king97} however, the {\it gf}-value used here has been increased by 0.110 dex.

\section{Conclusions}
Via a combination of small changes to the {\it gf}-values of a number of lines across the spectral region synthesised and a few more major changes in the immediate vicinity of the Be {\sc ii} doublet lines an accurate fit to the spectra of both the Sun and Procyon have been achieved.

Furthermore, the same line list is capable of synthesising the high resolution VLT/UVES spectra of four stars in NGC6633 with temperatures ranging from 6300K to 7600K \citep{ashwell05}.

\begin{acknowledgements}
JFA acknowledges the financial support of the UK Particle Physics and Astronomy Research Council (PPARC).
\end{acknowledgements}

\bibliographystyle{aa}

\begin{thebibliography}{}

\bibitem[Anders \& Grevesse(1989)]{anders89} Anders, E., \&
Grevesse, N.\ 1989, \gca, 53, 197

\bibitem[Ashwell et al.(2005)]{ashwell05} Ashwell, J.F., Jeffries, R.D., Smalley, B., Deliyannis, C.P., Steinhauer, A., \& King, J.R.\ 2005, accepted by MNRAS

\bibitem[Bagnulo et al.(2003)]{bagnulo03} Bagnulo, S., Jehin, E.,
Ledoux, C., Cabanac, R., Melo, C., Gilmozzi, R., \& The ESO Paranal Science
Operations Team 2003, The Messenger, 114, 10

\bibitem[King et al.(1997)]{king97} King, J.~R., Deliyannis,
C.~P., \& Boesgaard, A.~M.\ 1997, \apj, 478, 778

\bibitem[Kurucz et al.(1984)]{kurucz84} Kurucz, R.~L., Furenlid,
I., \& Brault, J.~T.~L.\ 1984, National Solar Observatory Atlas, Sunspot,
New Mexico: National Solar Observatory, 1984,

\bibitem[Lemke et al.(1993)]{lemke93} Lemke, M., Lambert,
D.~L., \& Edvardsson, B.\ 1993, \pasp, 105, 468

\bibitem[Sneden(2002)]{sneden02} Sneden, C.\ 2002, {\sc moog} An LTE Stellar Line Analysis Program

\end{thebibliography}

\end{document}